\documentclass[fleqn,10pt,twocolumn]{wlscirep}
\pdfoutput=1
\usepackage[utf8]{inputenc}
\usepackage[T1]{fontenc}
\usepackage{wrapfig}
\usepackage[separate-uncertainty=true]{siunitx} 
\title{Deterministic Single Ion Implantation with 99.87\% Confidence for Scalable Donor-Qubit Arrays in Silicon}

\author[1,2]{Alexander M. Jakob}
\author[1,2]{Simon G. Robson}
\author[1,3]{Vivien Schmitt}
\author[1,3]{Vincent Mourik}
\author[4]{Matthias Posselt}
\author[2,5]{Daniel Spemann}
\author[1,2]{Brett C. Johnson}
\author[1,3]{Hannes R. Firgau}
\author[6]{Edwin Mayes}
\author[1,2]{Jeffrey C. McCallum}
\author[1,3]{Andrea Morello}
\author[1,2,*]{David N. Jamieson}
\affil[1]{ARC Centre for Quantum Computation and Communication Technology (CQC$^2$T)}
\affil[2]{School of Physics, University of Melbourne, Parkville, 3010, VIC, Australia}
\affil[3]{School of Electrical Engineering and Telecommunications, UNSW Sydney, Sydney, 2052, NSW, Australia}
\affil[4]{Helmholtz-Zentrum Dresden-Rossendorf (HZDR), Dresden, 01328, Saxony, Germany}
\affil[5]{Leibniz Institute of Surface Engineering (IOM), Leipzig, 04318, Saxony, Germany}
\affil[6]{RMIT Microscopy and Microanalysis Facility, RMIT University, Melbourne, 3001, VIC, Australia}

\affil[*]{d.jamieson@unimelb.edu.au}

\begin{abstract}
The attributes of group-V-donor spins implanted in an isotopically purified $^{28}$Si crystal make them attractive qubits for large-scale quantum computer devices. Important features include long nuclear and electron spin lifetimes of $^{31}$P, hyperfine clock transitions in $^{209}$Bi and electrically controllable $^{123}$Sb nuclear spins. However, architectures for scalable quantum devices require the ability to fabricate deterministic arrays of individual donor atoms, placed with sufficient precision to enable high-fidelity quantum operations. Here we employ on-chip electrodes with charge-sensitive electronics to demonstrate the implantation of single low-energy (14~keV) P$^+$ ions with an unprecedented $99.87\pm0.02$\% confidence, while operating close to room-temperature. This permits integration with an atomic force microscope equipped with a scanning-probe ion aperture to address the critical issue of directing the implanted ions to precise locations. These results show that deterministic single-ion implantation can be a viable pathway for manufacturing large-scale donor arrays for quantum computation and other applications.
\end{abstract}
\begin{document}

\flushbottom
\maketitle
%
%
\thispagestyle{empty}

\section*{Introduction}
The development of quantum computers has reached the stage where noisy, intermediate-scale quantum (NISQ) \cite{preskill2018} devices with $\sim 50 - 100$ qubits can surpass classical supercomputers in executing some specific algorithms \cite{arute2019}. Even at the NISQ stage, the error budgets for the physical qubits are strict, requiring errors well below 1\% in order to achieve sufficient circuit depths. Beyond NISQ, error-corrected, universal quantum processors of the kind necessary to run Shor's factoring algorithm on a 2,000 bit classical key will require upwards of 4,000 logical qubits. Using a 2-dimensional surface code architecture, this would translate to about 200 million physical qubits with present error rates of around 0.1\% \cite{Fowler2012}. Future devices with lower error rates will reduce the required number of physical qubits. The surface code is also able to tolerate $5 - 10$\% physically non-functional (absent or faulty) qubits in the architecture \cite{Nagayama2017, PhysRevA.96.042316}.\\ 
Taking these constraints into account, a scalable universal quantum computing platform requires: (i) manufacturability at the $\sim 10^9$ physical qubit scale; (ii) physical gate error rates at or below 0.1\%; (iii) no more than a few percent of faulty qubits. Leading technologies including superconducting qubits and ion traps satisfy requirements (ii) and (iii). However, requirement (i) appears extremely challenging for these technologies where the physical qubits are spaced on a scale of several microns.\\   
Classical silicon devices can be manufactured using industry-standard metal-oxide-semiconductor (MOS) methods that yield billions of transistors on a $\sim 30$~nm pitch \cite{fischer2015}; their extension to quantum devices can thus naturally address requirement (i). This has motivated the development of silicon spin qubits \cite{zwanenburg2013}, starting from the donor-based proposal of Kane \cite{kane1998}. The electron \cite{Pla2012} and the nuclear \cite{Pla2013} spin of a single $^{31}$P donor, ion-implanted in a silicon MOS device, have proven to be outstanding qubits, with coherence times exceeding 0.5~s (electron) or 30~s (nucleus) \cite{muhonen2014}. Single-qubit error rates are in the 0.01\% - 0.03\% range \cite{Muhonen2015,dehollain2016}, thus addressing requirement (ii) at the 1-qubit level. Conditional two-qubit operations between exchange-coupled donors have been recently demonstrated \cite{madzik2020}.
In this work, we provide the first experimental evidence that the implantation of individual dopants can be detected with such high confidence to not constitute a barrier to the fulfilment of requirement (iii), i.e. a low density of faulty or absent qubits.\\
All examples of coherent quantum control of single-donor spin qubits in silicon have been so far obtained in devices where a small number of donors, subject to Poisson statistics, were introduced in the chip by ion implantation \cite{Jamieson2005,vanDonkelaar2015}. This follows the well-established precedent of ion implantation to introduce dopants in classical MOS devices \cite{rubin2003}. However, for the goal of manufacturing a large-scale quantum computer with a billion-qubit array of controllable donors in silicon, it will be essential to precisely and deterministically place the individual donors within the array.\\
A key benefit of ion implantation is that all group-V donors can be introduced into the silicon, allowing a diverse range of applications. $^{31}$P is the simplest system, offering spin-1/2 nuclear and electron spin qubits \cite{Pla2012,Pla2013,muhonen2014}. $^{123}$Sb has a nuclear spin 7/2 which can encode error protected logical qubits \cite{gross2020} and can be controlled by local electric fields \cite{asaad2020}. $^{209}$Bi has a large electron-nuclear hyperfine coupling that results in the formation of noise-protected ``clock transitions'' \cite{wolfowicz2013}. Utilizing these donors for quantum information requires placing them $\sim 20$~nm under the surface, so they can be addressed and read out with suitable nanoelectronic circuitry. As a consequence, the kinetic ion implantation energy lies in the range of $\sim  8-35$~keV. Achieving both, deterministic ion implantation of individual donors at such low energy, and localisation of each implant to high spatial precision, represents an ongoing challenge.

Several alternative strategies that address this challenge are in an advanced stage of development. The cold-ion trap \cite{Meijer2006,Schnitzler2009,Groot2019} and the fly-by image charge detector \cite{Raecke2019} are both deterministic ion source concepts, where the incidence of a single ion is detected  prior to implantation. These approaches do not impose any special requirements on the substrate and can therefore be used for many materials as well as silicon.\\
An earlier approach, analogous to Scanning Electron Microscopy (SEM), employs the ion-impact-induced burst of secondary electrons escaping from the substrate surface to count dopant atoms implanted into silicon devices \cite{Shinada_1999,Schenkel2002,Shinada2005}. However, the yield of secondary electrons is typically below 10~e$^{-}$/ion for the implant energies of interest here \cite{Wang2007, Xu2012} which limits the single ion detection confidence with conventional secondary electron detectors to $\approx90$\% \cite{Shinada_1999}.\\
In this work we adopt a method that detects the electron-hole (e-h) pairs generated by an ion impact in a silicon substrate by utilising on-chip detector electrodes. The on-chip electrodes form a reverse-biased p-i-n diode, as developed in solid-state detector technologies for ionizing radiation \cite{vanDonkelaar2015, Jamieson2016}. This method, based on the Ion Beam Induced Charge (IBIC) principle \cite{Breese1992}, is well established for high-energy (of order MeV) ions, but demonstrated here with keV ions. Thanks to a typical $\sim 1000$ e-h pairs produced by the ion impact, this method has the potential to provide high-confidence signals but, until now, a rigorous quantitative assessment of such confidence was still lacking.  

Furthermore, here we integrate single ion detectors and charge-sensitive electronics (Fig.~\ref{fig:Fig1}a) with an Atomic Force Microscope (AFM) nanostencil scanner (Fig.~\ref{fig:Fig1}b) for precision localisation of the implant site \cite{Persaud2005,Meijer2008}. We address some of the challenges of using this system to build, e.g., a donor-spin qubit architecture that utilises flip-flop qubits \cite{Tosi2017}. These are typically placed on a two-dimensional array with 200~nm pitch and coupled by electric dipole interactions (Fig.~\ref{fig:Fig1}c). Each donor must be located at a shallow depth, $\sim 7-25$~nm beneath a thin gate oxide so that it can be tunnel-coupled to a readout device \cite{Morello2010} and electrostatically controlled by metallic surface gates.\\ 
We further demonstrate that the ion impact signal can also be used to assess the physical characteristics of the ion stopping trajectory, which is subject to random collision events called straggling. Statistically rare events that result in an undesirable ion placement location can be identified from the signal characteristics. This unique capability distinguishes our IBIC principle from all other deterministic implantation approaches. Suitable algorithms, capable of signal pulse shape discrimination, could increase the yield of functioning donor qubits in ultra-scaled dopant arrays by employing active correction protocols such as conditional implant-repetition steps and dynamic array reconfiguration.\\ 
However, to exploit this capability requires charge-sensitive signal processing electronics for the $\sim 1,000$ e-h pairs typically generated by each ion impact for shallow implantation. Cryogenic operation of the substrate and electronics is commonly applied to achieve sufficiently low noise thresholds. However, cryogenic systems are not readily compatible with the integration into ancillary apparatus that must operate at room temperature and can impose considerable operation complexity \cite{Singh2016}.\\
Here, we present a reliable, high-fidelity, counted single-ion implantation system operating near room temperature. We accurately benchmark the noise and error budget of the system and extract a detection confidence approaching 99.9\% for $^{31}$P$^+$ ions implanted at 14~keV. This system is compatible with subsequent processing steps required to fabricate multi-qubit devices.

\section*{Single-Ion Implant Detection}
\begin{figure}[t!]
\centering
\includegraphics[width=0.98\linewidth]{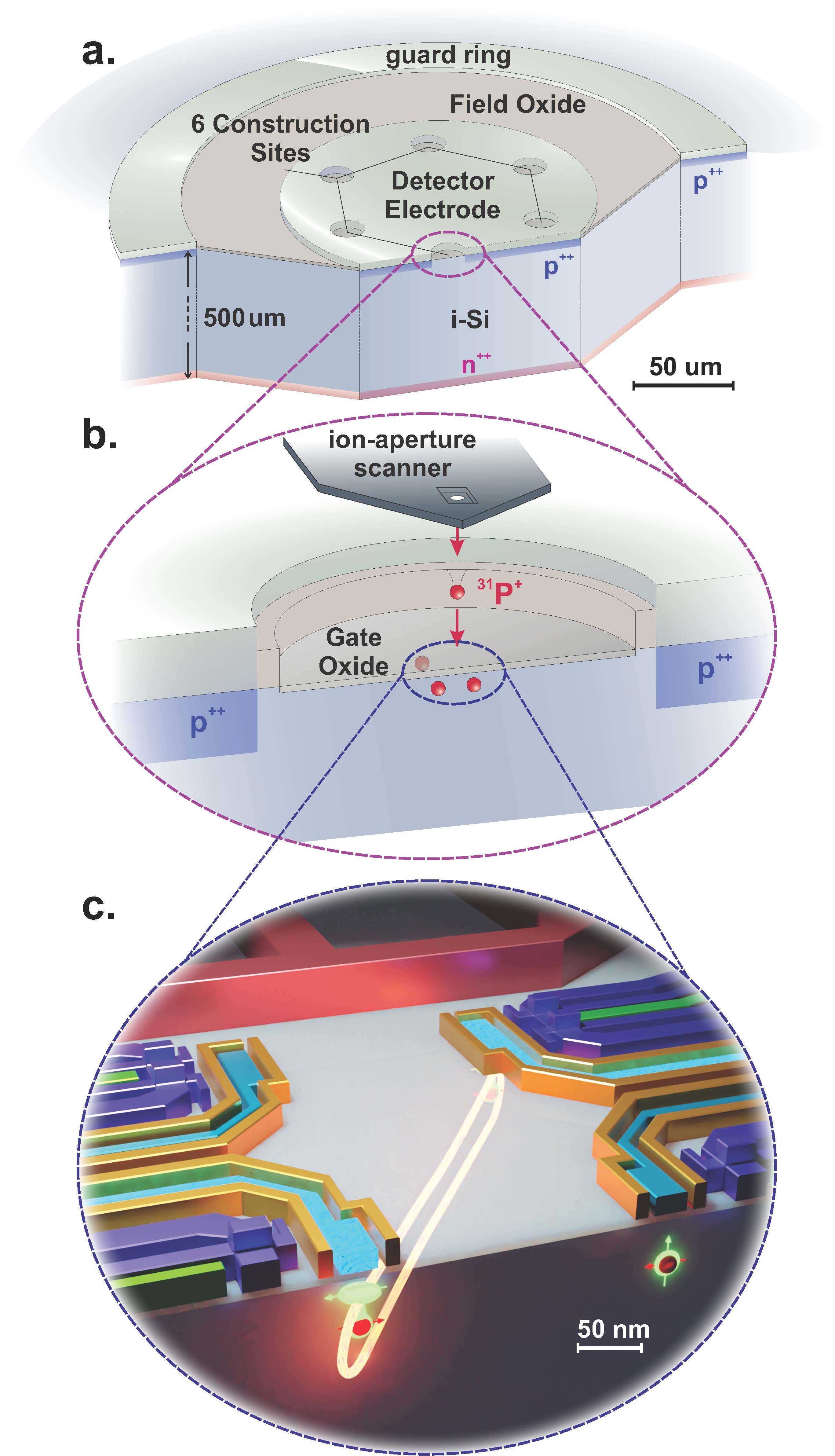}
\caption{\textbf{Localisation of single ion implants} \textbf{a}. Schematic of the silicon single ion detector die, incorporating a vertical `sandwich-type' p-i-n detector geometry. The detector incorporates an inner circular top electrode and an outer grounded p-type guard ring to minimise leakage current. The inner top electrode comprises six circular construction sites each with a uniform 5~nm SiO$_2$ thick gate oxide above intrinsic (100) silicon. \textbf{b}. Formation of a donor array by deterministic step-and-repeat single ion implantation in the selected construction site. The AFM cantilever, which incorporates a nanostencil aperture, acts as a movable mask for the ion beam. The signal from a single ion implant event triggers the AFM nanostencil scanner to step to the next implant site. 
\textbf{c}. Schematic of a $2 \times 2$ $^{31}$P-donor array with $\approx 200$~nm spacing, as appropriate for flip-flop qubit devices \cite{Tosi2017}. These qubits employ long-range electric dipole interactions, so that entangling gate operations can be performed even beyond the nearest-neighbours, for instance across the diagonal of the array. The control and readout circuitry is fabricated after the implantation and the rapid thermal anneal for donor activation. }
\label{fig:Fig1}
\end{figure}
The devices presented here employ a substrate configured with multiple construction sites, which will allow us to fabricate multiple single- or few-qubit devices on a single chip. Each construction site has a lateral diameter of \SI{15}{\micro\meter}. As shown in Figure~\ref{fig:Fig1}a, the construction sites are surrounded by a detector top electrode, and each site features a pre-fabricated thin gate oxide needed for subsequent integration of qubit control nano circuitry. The top electrode makes contact with a boron-doped p-well on the intrinsic silicon substrate, with a n-type back contact forming the p-i-n detector. To meet the low noise performance requirements, two important design features are employed: (i) A grounded p-type guard ring surrounds the top electrode and screens the active detector volume against parasitic free charge carriers from outer interface and bulk defects \cite{Evensen1993} and minimises the reverse bias leakage current, which would otherwise obscure ion impact signals; (ii) Minimising the top electrode area lowers the total device capacitance and consequently the parallel white noise contribution in the charge-sensitive preamplifier.\\
The principle of controlled donor array formation inside a selected construction site is illustrated in Fig.~\ref{fig:Fig1}b. An AFM nanostencil scanner localises the implant site to high spatial precision and steps to the next array site when  triggered by the detected ion implant signal. The signal can also trigger a fast ion beam blanker (typically $\sim 100$~ns response time) to minimise the probability of further implant events at the same array site.\\   
The gate dielectric is a ~$5$~nm thin high-quality SiO$_2$ oxide, thermally grown (see Methods) in advance of all other fabrication steps, because the required thermal budget is not compatible with subsequent fabrication steps. Moreover, the thermal growth of the gate oxide has the advantage of passivating interface charge traps and reducing fixed oxide charges that would otherwise reduce the ion-induced charge signal in the detector. The ions traversing the gate oxide during implantation suffer from some kinetic energy loss that is not available for the signal generation. However, this effect is tolerable given the low oxide thickness and the excellent signal detection efficiency enabled by this surface passivation.\\ 
The critical properties of the gate oxide in the present detectors are measured from MOSCap devices processed together with the detector wafers. They amount to $\leq6\times10^{9}$~eV$^{-1}$cm$^{-2}$ for the fixed oxide charge density and $\leq8\times10^{10}$~cm$^{-2}$ for the oxide interface trap density. These values are found to be sufficiently low to ensure signals close to 100\% of the charge created by single ion implant events at implantation energies of interest. In a broader context, these values also indicate that the devices have a sufficiently low density of charge defects for high-fidelity operation of the donor spin qubits that will result from this fabrication process.\\

\section*{Single-Ion Implant Localisation}
\begin{figure}[t!]
\centering
\includegraphics[width=0.98\linewidth]{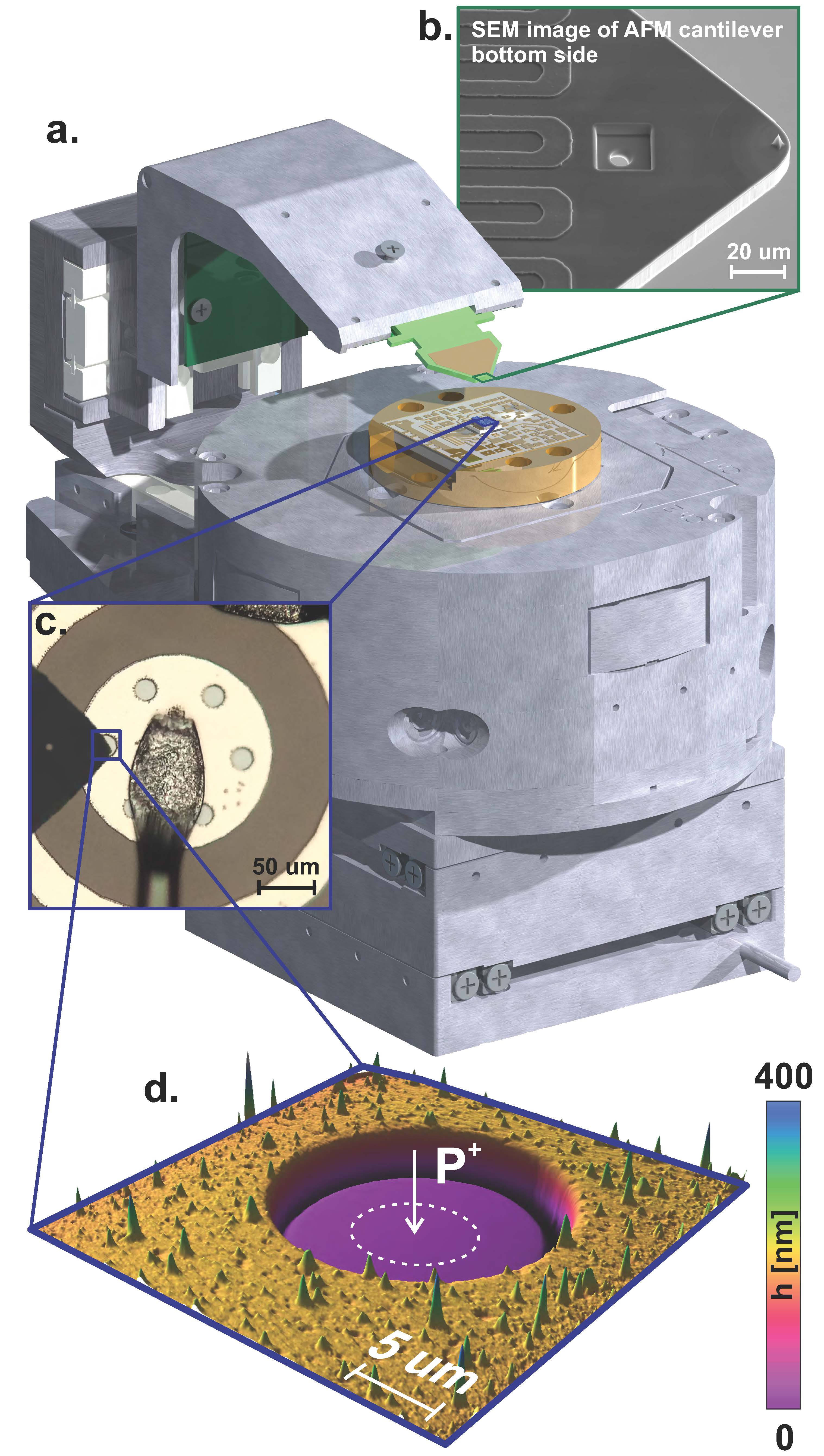}
\caption{\textbf{The single ion implanter system} \textbf{a.} The system incorporates an AFM, where the integrated sample stage also houses the charge-sensitive preamplifier electronics \cite{Bertuccio1993}. The stage incorporates Peltier cooling to 263 K. \textbf{b.} SEM micrograph of the AFM cantilever (see text). For the present study we equipped the cantilever with a micro-aperture of $8 \times 8$~$\mu$m$^2$, sufficient to localise the ion beam within a single construction site. \textbf{c.} In-situ optical camera view of the detector, showing the cantilever and the wire bond connecting the detector top electrode to the charge-sensitive preamplifier circuit board housed within the sample stage. \textbf{d.} The AFM image from \textbf{c} showing the selected construction site as a  3D topography map. The irradiated region localised by the nanostencil is highlighted by the dashed circle. Lithographic alignment markers mapped by the AFM (not shown) allow nanometer precision alignment between ion implant sites and subsequent processing steps.}
\label{fig:Fig2}
\end{figure}

The single ion implantation detection system operates in conjunction with an AFM \cite{nanoanalytik2020} that is equipped with a nanostencil integrated into the cantilever, as shown in Figure~\ref{fig:Fig2}. The sample stage of the AFM holds the substrate chip to be implanted and also incorporates the charge-sensitive electronics coupled to the on-chip detectors. The AFM cantilever is controlled by integrated self-actuating technology \cite{Majstrzyk2018} instead of conventional laser-based optical schemes, which are incompatible with the on-chip detectors as they would be swamped by light spillage. Additional benefits include a more compact and sturdy AFM design that is less sensitive to thermo-mechanical drift. The AFM cantilever employs tapping mode \cite{Binnig1986} to approach and image the substrate. This approach is benign compared to more invasive techniques such as electron microscopy, which could inject excess charge and degrade the gate oxide passivation.\\ 
The AFM cantilever nanostencil assembly incorporates a Focused Ion Beam(FIB)-milled \cite{Watkins1986} collimator for ion-implantation. In the results presented here we employed an 8~$\mu$m diameter aperture because we sought to investigate the physics of the ion-solid interaction from ion impact signals that are randomly distributed over one construction site while avoiding edge effects. Sub-10~nm ion apertures are readily available for controlled donor array formation experiments, and will be described in a separate study. The AFM cantilever is operated by a  monolithic top stage with an independently controllable travel range of $18 \times 18 \times 8$~mm in all three dimensions. This top stage allows a rapid and precise alignment of the cantilever aperture with respect to the incident ion micro-beam, which has $\approx 20$ $\mu$m diameter. The much larger cantilever dimension of $350\times120$~$\mu$m$^{2}$ ensures that no unintentional ion strikes occur outside the collimator.\\
The device for implantation is mounted on a circuit board that also contains the charge-sensitive preamplifier electronics. This assembly is mounted on the AFM stage within a Faraday shield containing a thermo-electric Peltier cooler for operation of the detector at 263~K (-10$^{\circ}$C). An opening in the shield enables access for the AFM cantilever and the stage provides a $60\times60$~$\mu$m$^{2}$ lateral travel range with nominally 5~$\textrm{Å}$ repeated positioning accuracy. 
The entire AFM assembly is mounted on a positioning stack with $\pm$15~mm lateral travel. This allows coarse positioning between sample and cantilever with nominally $\sim 50$~nm placement accuracy.  An AFM image of a construction site is shown in Fig.~\ref{fig:Fig2}c, which shows the required uniformity of the surface needed for ion implantation of the near-surface donors. The AFM micrograph can also identify location markers (not shown) to align the cantilever ion aperture with the required implant sites to high precision.\\

\section*{Induction and Detection of Charge Signals}
Detecting with high confidence the signal from $\lesssim 1000$ e-h pairs induced by a single ion implant requires a minimisation of the noise generated by the detector and the associated electronics. The noise performance of a solid-state detector is mainly determined by the combined leakage current $I_{\mathrm{tot}}$ and capacitance $C_{\mathrm{tot}}$ of the detector and its first-stage amplifier. For ultra-low noise applications, the latter typically consists of a junction field effect transistor (JFET), whose internal gate design and fabrication technology \cite{Betta1998, Betta2001} determine $I_{\mathrm{tot}}$ and $C_{\mathrm{tot}}$.\\ 
\begin{figure}[t!]
\centering
\includegraphics[width=0.98\linewidth]{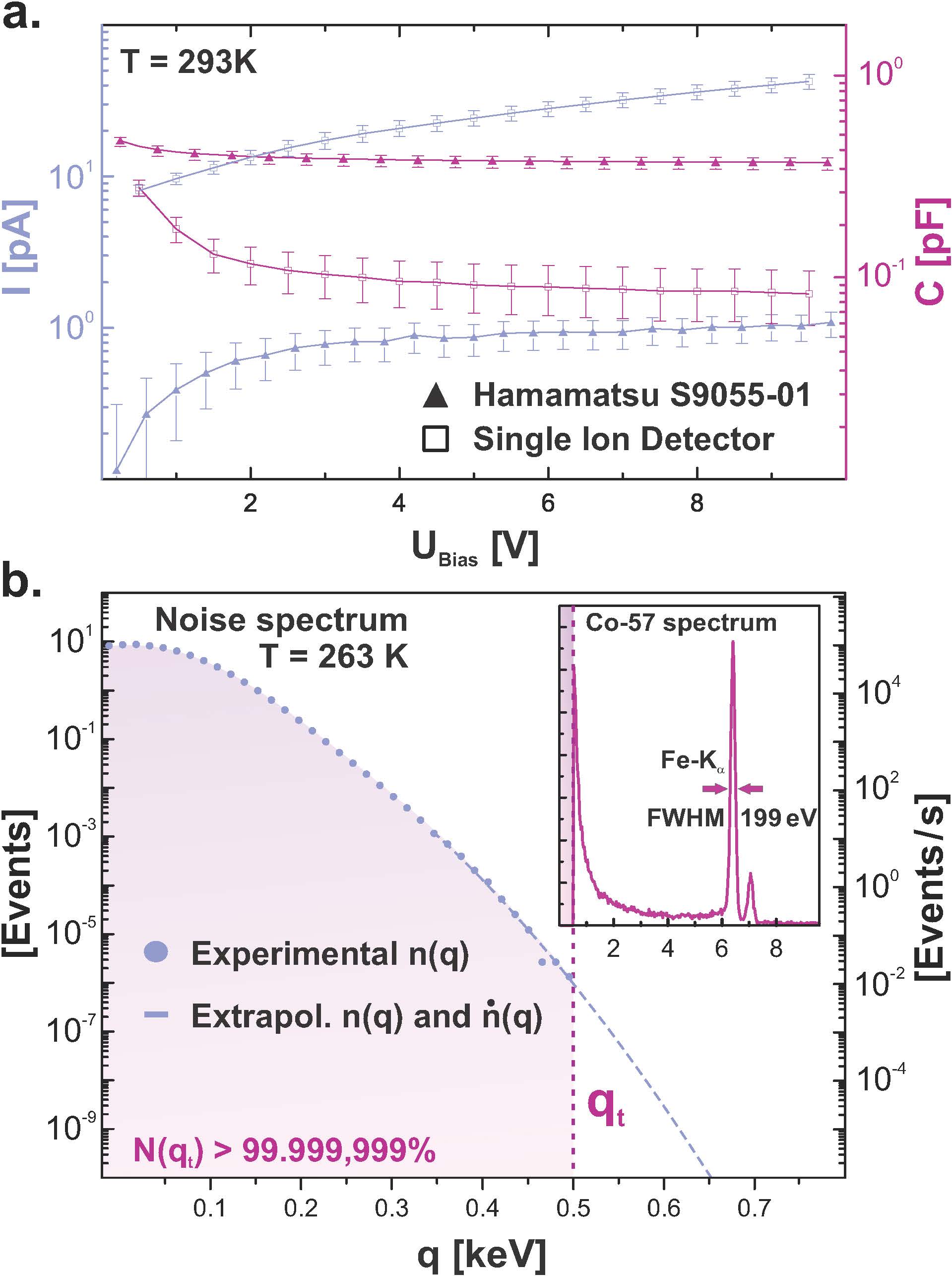}
\caption{\textbf{Detector characteristics} \textbf{a.} I-V and C-V data of the detector die compared to a reference photodiode. The detector has a comparable low capacitance but a leakage current an order of magnitude higher, suggesting further improvements are possible. \textbf{b.} The detector dark noise spectrum $n$ and related rate $\Dot{n}$ as functions of the charge-equivalent signal amplitude $q$. A Rayleigh distribution is used to model the noise spectrum. A threshold  $q_\mathrm{t}$ is delineated here where 99.999,999\% of the cumulative noise $N(q_\mathrm{t})$ (pink area) is discarded by a lower-level discriminator in the data acquisition electronics. The inset shows the $^{55}$Fe X-ray K-emission lines of a $^{57}$Co radionuclide, acquired with all system components activated, including the AFM system. A resolution of about 200~eV FWHM for the 6.4~keV K$_\mathrm{\alpha}$ peak acquired at a substrate temperature of 263~K verifies negligible signal degradation from cross-talk from the AFM system. This signal spectrum $s(q)$ also allows a quantitative calibration of the charge-equivalent signal axis $q$ in units of [keV].
}
\label{fig:Fig3}
\end{figure}

Our system incorporates several optimisation strategies developed for different applications.  State-of-the-art radiation detectors for X-rays feature an integrated JFET and exhibit a capacitance on the order of $\mathit{C}_{\mathrm{tot}} \leq 300$~fF \cite{Fiorini2005}. Fast-recovery p-i-n photo diodes can have a full-depletion reversed bias leakage current as low as $\mathit{I}_{\mathrm{tot}}\leq$~1~pA at room temperature \cite{hama2020}. These highly application-optimised detectors provide a benchmark for the performance of our devices optimised for deterministic doping.\\ 
Figure ~\ref{fig:Fig3}a illustrates capacitance-voltage (C-V) and current-voltage (I-V) graphs representative of the present device alongside results from a reference photodiode \cite{hama2020}. 
By minimising the top electrode area that contains the construction sites, and incorporating a p-guard ring surrounding the top electrode to suppress leakage current, we obtained a capacitance of ~$80\pm30$~fF (a factor four lower than the reference diode) and a leakage current of approximately $35\pm10$~pA at 10~V reverse bias and room temperature operation. Moderate cooling to -10$^{\circ}$C gives the required sub-pA leakage current. Further improvement of these values is possible by optimisation of the detector fabrication process.\\ 
Experimental measurements of the detector noise are obtained using a $^{57}$Co radionuclide, emitting characteristic Fe K$_{\alpha}$ and K$_{\beta}$ X-ray photons at 6.40~keV and 7.06~keV respectively. Upon absorption inside the active detector volume, an X-ray photon excites a  number of e-h pairs proportional to its energy. Drift of the e-h pairs in the bias field results in a charge-equivalent voltage pulse $q$ at the detector electrodes whose amplitude is in turn proportional to the photon energy $ E_\mathrm{X-ray}$ and commonly expressed in units of keV. The signal spectrum, $s_\mathrm{X-ray}(q)$, also features Bremsstrahlung and detector noise.
A representative X-ray spectrum from a present detector, operated at 263~K, is shown in Fig.~\ref{fig:Fig3}b. The closely spaced  Fe-K$_{\alpha}$ and K$_{\beta}$ photon energies are clearly resolved. The K$_{\alpha}$-peak exhibits an FWHM of about 200~eV, which corresponds to an r.m.s. noise of $\sigma_{\mathrm{noise}} \approx 70$~eV and thus comparable to state-of the-art silicon detectors for other applications. The detector dark noise spectrum, $n(q)$, obtained with no radionuclide present is shown in Figure~\ref{fig:Fig3}b. As demonstrated in the following, the steep noise attenuation as a function of energy of $\approx 2.3$~dB per 100~eV is essential for low energy single-ion detection with high confidence.\\ 

\section*{Deterministic Ion Implantation}
The stopping of an ion as it dissipates kinetic energy in a crystal is caused by nuclear and electronic energy loss \cite{Biersack1980, Ziegler2010} in proportions that depend on the ion mass and energy. The electronic stopping fraction $\mathit{f_{\mathrm{el,}i}}$ of a single ion generates e-h pairs and thus the energy-equivalent signal amplitude $q_i$ utilised for the event detection. For an ensemble $\{i\}$ of consecutive ion implants, each event $i$ contributes to a signal spectrum $s(q)=\{q_i\}$. Electron-hole (e-h) pairs generated within the gate oxide (i.e. outside the silicon crystal) and losses from charge recombination reduce the signal amplitude $q_i$. Further effects like ion channeling and substrate atom recoils influence the electronic fraction $\mathit{f_{\mathrm{el,}i}}$ itself. The interaction of these factors leads to characteristic high and low energy tails in the spectrum $s(q)$, which consequently reflects both the stopping physics specific for an ion species as well as the detector properties. 
\begin{figure*}[t!]
\centering
\includegraphics[width=0.98\linewidth]{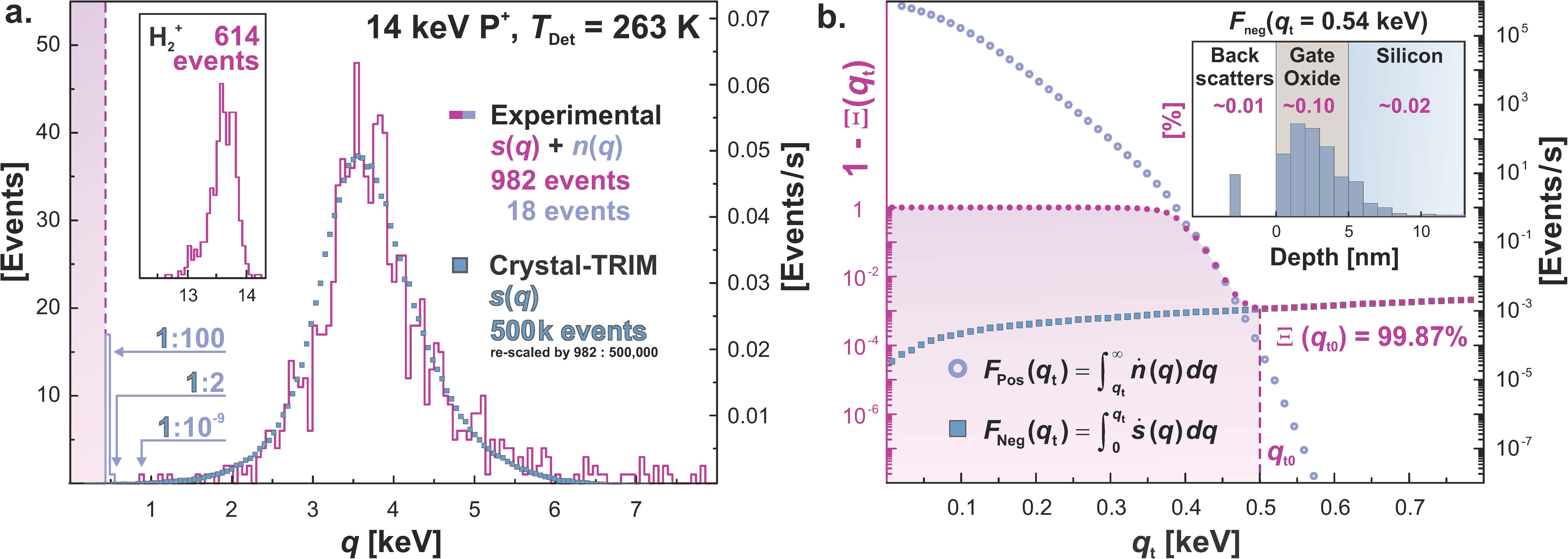}
\caption{\textbf{Counted ion implantation - experiments and analysis} \textbf{a.} Experimental signal spectrum $s(q)+n(q)$ for 14~keV P$^{+}$ ions implanted inside the construction site of a single ion detector. The spectrum contains 1000 events in total, with 18 events assigned to noise $n(q)$ (light blue) and remaining 982 events $s(q)$ (pink) induced by ions with a residual uncertainty below 10$^{-9}$. The related time-derivatives (rates) $\dot{s}(q)$ and $\dot{n}(q)$ can be determined with the known total acquisition time of 760~s. A numerical simulation with optimised parameters (see text) of $s(q)$ for 500,000 virtual implants (blue square scatter plot) is overlaid on the experimental spectrum and demonstrates close agreement. The experimental spectrum for 614 14~keV H$_{2}^{+}$ molecule-ion implants appears in the inset. These ions produce signals well above the noise threshold and confirm 100\% charge collection efficiency of the detector. \textbf{b.} The cumulative "false positives" $F_\mathrm{Pos}$ and "false negatives" $F_\mathrm{Neg}$ are plotted alongside the ion ensemble detection diffidence $1-\varXi$ as a function of the discriminator threshold level $q_\mathrm{t}$. The optimum value $q_\mathrm{t0}$ yields a nominal detection confidence of 99.87\%. The inset depicts the final calculated ion placement depths of the sub-threshold signal events $F_\mathrm{Neg}(q_\mathrm{t0})$ below $q_\mathrm{t}$. The majority consists of non-critical ion backscatter events and ions stopping inside the gate oxide.}
\label{fig:Fig4}
\end{figure*}

To measure the response of the present detector to ion implantation, the AFM cantilever with an  8~$\mu$m diameter aperture was used to localize the ion beam to the construction sites. We first show a spectrum from 14~keV H$_{2}^{+}$ ions. This molecular ion dissociates instantaneously upon impacting the surface gate oxide, yielding two 7~keV H$^{+}$ ions, each having $\sim 100$~nm average penetration depth.  The e-h pairs produced by both ions produce a combined signal pulse in the detector.  The low mass of protons results in $\sim$ 95\% of their kinetic energy dissipating in e-h-pair generation \cite{Biersack1980, Ziegler2010}. The signal spectrum (Fig.~\ref{fig:Fig4}a inset) $s(q)$ thus exhibits a sharp peak at an energy very close to the 14~keV energy of the incoming H$_{2}^{+}$ ions. This result demonstrates the expected detector performance and minimal losses due to charge trapping and recombination. Furthermore, essentially no low energy signal events are registered outside the main peak body, indicating negligible scattering artifacts from the ion aperture.\\ 
The construction site was subsequently irradiated with 14~keV P$^{+}$ ions for approximately 760~s and a beam current of $\sim 80$~ions/min. The corresponding signal spectrum consists of $\sim 1000$ detected events, as illustrated in Fig.~\ref{fig:Fig4}a. Note that a $2\times2$ flip-flop qubit prototype device as per Fig.~\ref{fig:Fig1}c requires an inter-qubit pitch of about 200~nm. This pitch leads to an average surface doping density of 25~atoms/$\mu$m$^{2}$, which is very similar to the density produced with the present implant parameters. It furthermore corresponds to a sufficiently low ion fluence to not cause any measurable charge collection efficiency deterioration from ion-induced substrate damage \cite{Auden2015}.\\ 
The intended application of our system is for near-surface doping, including the 14~keV P$^+$-ion implants presented here. The corresponding spectrum, shown in Fig.~\ref{fig:Fig4}a, exhibits a peak at 3.6~keV, which is significantly less than the incident ion kinetic energy, owing to the smaller fraction of electronic energy loss to generate e-h pairs compared to lighter ions. Also, the straggling from statistical variations in the ion stopping process leads to a more pronounced variation in the fraction of electronic energy loss per ion compared to 7~keV H$^{+}$ ions. Consequently, the spectrum is broader and skewed towards low energy signals. Ion channeling, mainly along the Si [100] axis, enhances electronic energy loss and leads to a high-energy tail in the signal spectrum. Recoiling Si and O atoms generated by P$^+$-ions within the oxide layer are responsible for the low-energy signals near the detection threshold $q_{\mathrm{t}}$.\\

\section*{Deterministic Implantation Confidence}
We now consider the implications of this experimental spectrum for the use of deterministic doping in the fabrication of large-scale donor arrays, e.g. for donor-based quantum computers. We estimate the detection confidence by separating the ion-induced signal spectrum $s(q)$ from the noise spectrum $n(q)$. This is achieved by combining a computational model with realistic experimental parameters. The trajectories of individual ions can be modeled to provide insights into the final location of the ion in the substrate and the corresponding signal amplitude from the detector.\\
We consider here a binary collision model for the ion-solid interaction to compute the associated electronic energy loss and hence the detector signal. The model first uses the TRIM code to compute the ion trajectory through the surface gate oxide and to determine ion position and velocity vector at the interface to the silicon substrate (including recoiling Si and O atoms). Then, a modified Crystal-TRIM code is used to compute the ion trajectory inside the (100) crystalline silicon and determine the total electronic energy loss of the ion and associated recoils. Details of the simulation procedure are explained in the Methods section. A set of semi-empirical parameters (see Methods section) are fitted to match the experimental signal spectrum.\\
Results from this model to compute the signal spectrum of 500,000 14~keV P$^+$-ions are shown in Figure~\ref{fig:Fig4}a. The simulation agrees with the experimental spectrum within Poisson statistics. A sparse set of signals visible in the experimental spectrum above 6~keV appears to point to physical processes not included in the model. The model assumes the surface gate oxide to be homogeneous, whereas the actual oxide has an amorphous structure leading to inevitable small variations in the density and thus the scattering dynamics, which are neglected by the model. 
Nevertheless, the satisfactory match, especially in the low-energy signal regime, justifies the use of this simulation procedure to assess the detection confidence for our experiments with 14~keV P$^+$-ions.\\
For a given experimental data acquisition time, the confidence that a signal arises from a single ion implantation event is limited by the probability that a noise event occurs within the same time window. This confidence is in proportion to the number of e-h pairs available from the ion impact. A discriminator threshold, $q_\mathrm{t}$, is used to discard most of the low-energy events which are dominated by noise signals. The key feature of the experimental spectrum is the near-absence of signals in the energy window above the discriminator threshold, which is a testament to the extremely low noise obtained by our system. To demonstrate the role of the discriminator threshold, the experimental spectrum in Fig.~\ref{fig:Fig4}a was obtained with a discriminator threshold set to $q_\mathrm{t}=0.42$~keV (rose-coloured area) so that some signals otherwise rejected by the optimum discriminator threshold (discussed below) are retained in the experimental spectrum of Fig.~\ref{fig:Fig4}a (light-blue events).  We now examine these signals in detail.\\
First, we consider the noise signals registered immediately above the discriminator threshold. The event rate $\dot{n}(q)$ for the noise spectrum can be obtained from the Rayleigh function in Fig.~\ref{fig:Fig3}b and compared with the corresponding ion signal rate $\dot{s}(q)$ extracted from the modelled signal spectrum in Fig.~\ref{fig:Fig4}a.\\ 
For the first two energy bins $q\in(0.42,0.48]$~keV (17 counts) and $q\in(0.48,0.54]$~keV (1 count) just above the discrimination threshold $q_\mathrm{t}$, the probability $\dot{s}(q):\dot{n}(q)$ that the events binned therein are from ions is about $1:100$ and $1:2$, respectively. In contrast, the single event registered at $q=0.9$~keV has an ion signal probability of better than $\gtrsim 1:10^{-9}$ and therefore has high confidence of being an ion implant event.\\ 
Consequently, the remaining 982 signal events binned beyond 0.9~keV are with near 100\% confidence due to ion implants because of the very low noise threshold of the detector. The residual uncertainty is mainly determined by rare environmental disturbances not considered here, such as power supply stability fluctuations or natural nuclear decays in the environment.\\
Second, we evaluate the detection confidence for ion implant signals which relies on the consideration of all relevant signals that are gathered or discarded by the detector system. Due to the partial overlap of the signal ($s(q)$) and noise ($n(q)$) distributions, it is possible to identify the optimum discriminator threshold level $q_\mathrm{t}$ by considering two critical quantities: false-positive signals from retaining noise above $q_\mathrm{t}$, and false-negative signals from discarding ion implantation events below $q_\mathrm{t}$. The false-positive rate is easily obtained by integrating the experimental noise spectrum $n(q)$ normalised to the acquisition time from $q_\mathrm{t}$ to $\infty$, i.e. $F_\mathrm{Pos}(q_\mathrm{t}):=\Dot{N}|_{q_\mathrm{t}}^{\infty}$. The false-negative rate $F_\mathrm{Neg}(q_\mathrm{t}):=\Dot{S}|_{0}^{q_\mathrm{t}}$ is derived via integration of the model signal spectrum $s(q)$ and a normalisation to the average ion rate $r_\mathrm{Ion}=\dot{S}|_{0}^{\infty}$.\\ 
Figure~\ref{fig:Fig4}b. illustrates the experimentally obtained $F_\mathrm{Pos}(q_\mathrm{t})$ event rate as well as the $F_\mathrm{Neg}(q_\mathrm{t})$ event rate for an ion rate of $r_\mathrm{Ion}=80\,\mathrm{min}^{-1}\approx1.3\,\mathrm{s}^{-1}$ (as adopted in the experiment reported in Fig.~\ref{fig:Fig4}a). \\ 
The ion detection confidence $\varXi$ (normalised as probability) can be then derived as (see Methods): 
\begin{equation}
\label{eqn:Xi}
\varXi(q_{\mathrm{t0}}):=\left(1-\frac{F_{\mathrm{Neg}}(q_{\mathrm{t0}})}{r_\mathrm{Ion}}\right)\,\left(\frac{r_\mathrm{Ion}}{r_\mathrm{Ion}+F_{\mathrm{Pos}}(q_{\mathrm{t0}})}\right)
\end{equation}
with the first factor describing the fraction of ions that create detectable signal events above the threshold level $q_\mathrm{t0}$ and the second factor stating the probability that the registered event was not due to noise. Equation~\ref{eqn:Xi} constitutes an optimisation problem with the threshold level $q_\mathrm{t0}$ adjusted to maximise $\varXi$. The ion detection diffidence $1-\varXi$ is illustrated in Fig.~\ref{fig:Fig4}b. For $q_\mathrm{t}$ set to a low threshold level, false positives $F_\mathrm{Pos}$ dominate the acquired signal and cause the diffidence to saturate close to 1. At the other extreme, for $q_\mathrm{t}$ set to a high threshold level, false negatives (rejected real ion implantation signals) $F_\mathrm{Neg}$ become the main confidence limitation. 
Noteworthy in Fig.~\ref{fig:Fig4}b is the shallow curve profile of $F_\mathrm{Neg}$ throughout the entire signal regime, causing $\varXi$ to remain below 99.99\% - regardless of the detector noise performance.\\
The optimum threshold level can be determined from the curves in Fig.~\ref{fig:Fig4}b to be $q_\mathrm{t0}\approx0.5$~keV with a nominal ion ensemble detection confidence of $\varXi(q_\mathrm{t0})= 99.87\pm0.02$\%. For the approximately 0.13\% implant events that produce signals below this threshold and only slightly above ($q_\mathrm{t}\lesssim0.54$~keV), our computational model allows an examination of the corresponding ion stopping trajectories. As tabulated in the inset of Fig.~\ref{fig:Fig4}b, the model shows that most of these ions ($0.11\%$) stop either inside the gate oxide or are backscattered at the sample surface. Only a residual $0.02\%$ of the 14~keV P$^+$-ions end up in the silicon substrate without producing a detectable signal. Hence, the majority of the sub-threshold signal events induced by ions are not detrimental in terms of qubit loss faults, because only ions reaching the silicon substrate form electrically active dopants.   

\section*{Conclusion}
We have presented a single-ion detector, integrated with an AFM nanostencil and operating near room-temperature, which allows deterministic ion implantation by detecting ion-induced e-h pairs inside the silicon substrate. The system can be employed with group-V dopant atoms, implanted near the surface of a silicon device, to form single-atom spin qubit devices such as the 2D architectures exploiting the flip-flop- qubit \cite{Tosi2017}. Our single-ion detector technology exhibits an exceptionally low noise background at near-room temperature, as shown with 14~keV P$^{+}$ ions. The system is compatible with many standard ion implanters, commonly equipped with stochastic ion sources that cause random ion arrival times at the substrate. The detector signals from implanted ions provide a characteristic spectrum that allows deeper insight into the ion-solid interaction. Thanks to an improved model, rare implantation events that produce sub-threshold ion signals could be investigated in detail.\\
For the configuration of our system, we conclude that the confidence of detecting a single ion implantation event takes the promising and unprecedented value of 99.87\%. Remarkably, the residual diffidence can be mainly attributed to the stopping physics of 14~keV P$^+$-ions in silicon, whereas the detector noise plays only a subordinate role here. 
Future studies will analyse advanced scalability projections for 2D qubit array formation and extend Equation \ref{eqn:Xi} to include e.g. the double-implant probability as a function of the incident ion beam current (see Methods section). These considerations will become important when seeking to increase the ion beam fluence to reduce the total implantation time for large donor arrays. Although the present study employed $^{31}$P donors, we expect comparable confidence levels for other dopant species, as long as the implantation energy is adjusted to preserve a similar number of $\sim 1000$ e-h pairs per ion impact.\\
In conclusion, our results show that a single 14~keV $^{31}$P-ion, implanted in a silicon device operated at near-room temperature and integrated with an AFM nanostencil scanner, can be detected with extremely high confidence. Therefore, ion detection uncertainties will not constitute an obstacle to the construction of a fault-tolerant, large-scale donor-based quantum computer in silicon.

\section*{Methods}

\textbf{Detector fabrication:} Standard metal-oxide-semiconductor (MOS) processing \cite{rossi2015} is employed to fabricate the detectors studied in this work. The initial wafer is a <100> Uniform High Purity Silicon (UHPS) wafer from Topsil. The low residual n-doping yields a resistivity of 9250 Ohm-cm. The on-chip single-ion detectors are fabricated as follow:
\begin{enumerate}
    \item Etching of alignment markers for subsequent optical lithographical steps: The pattern is first defined using standard optical lithography and then transferred into a previously grown wet thermal SiO$_2$ oxide. Using tetramethylammonium hydroxide (TMAH), the pattern is then etched into the silicon; the oxide is subsequently removed using BHF.
    \item Creation of the detector’s p-doped regions: The area to be doped is defined using standard optical lithography and then transferred onto a thermally-grown wet oxide. P-type doping is obtained by thermal diffusion of boron. Lastly, the oxide is removed using BHF.
    \item Back n-doping of the detector: First, a thermal oxide is grown in a steam ambient. Using photoresist as a mask, the oxide on the back of the detector is then removed using BHF. This then serves as the mask for the thermal diffusion doping of the rear of the detectors using phosphorous. The masking oxide is then removed using BHF.
    \item Growth of the thick field oxide using a dry oxidation process.
    \item Growth of the thin gate oxide using a dry oxidation process, in areas defined within the field oxide using optical lithography and etched using BHF.
    \item Etching of vias for metallisation using a photolithographic mask and BHF.
    \item Deposition of metallisation: A mask is first defined using photolithography and any native oxide that has grown onto exposed silicon is removed using a quick hydrofluoric acid dip. Using e-beam evaporation, 100~nm of aluminium is then deposited onto the wafer, directly followed by 10~nm of Platinum, both on the front and back. Metallisation in undesired regions is removed using a lift-off process using warm N-Methyl-2-pyrrolidone (NMP).
    \item Annealing of the detectors in a forming gas ambient (5\% hydrogen, 95\% nitrogen) at \SI{400}{\degreeCelsius}.
\end{enumerate}
\textbf{I-V/C-V analysis:} Detector chips are mounted on a chip carrier and placed inside a light-shielded analysis chamber held under rough vacuum ($\sim 1\times 10^{-3}$ Torr). The chip carrier is attached to a low-noise multiplexed feedthrough that allows interconnection to either a Keithley 6487 picoammeter (resolution 20~fA) or Boonton 7200 capacitance meter (resolution 1~fF) for I-V and C-V measurements, respectively.\\
\textbf{FOC/DIT analysis:} The Si/SiO$_2$ interface trap densities (Dit) were estimated using the Hill-Coleman method and deep level transient spectroscopy. Mid-gap values of test devices with a 5~nm oxide are found to be in the low $10^{10} \;\rm cm^{-2} eV^{-1}$ range. The fixed oxide charge was estimated by identifying the flat band voltage shift in a CV curve as a function of oxide thickness. Values are in the low $10^{11} \;\rm cm^{-2}$ range\\
\textbf{Charge-sensitive electronics:} The charge-sensitive preamplifier is based on the forward-biased FET circuit design of Bertuccio et al. \cite{Bertuccio1993}. The preamplified signal pulse is fed into an Amptek PX5 digital pulse processor, which performs trapezoidal pulse shaping ($\tau_\mathrm{peak}=9.6\mu$s) and multi-channel analysis to provide the final signal histogram data via USB connection to the Amptek MCA control- and acquisition software.\\
Additionally, the digital signal resolution was set to about 60~eV/channel (256 channels ranging from 0 to 15.4 keV) to give about 10\% statistical variation in the maximum counts per channel for the total fluence of 1000 ion counts. This choice is also consistent with the ion energy straggling which does not justify a higher resolution.\\
\textbf{Nanostencil fabrication:} The fabrication of the aperture in the AFM cantilever to form the nanostencil is done with an FEI Scios SEM/FIB system. The aperture can be made with a diameter from microns down to sub-10~nm via in-situ monitoring with the SEM.  After milling the aperture a Pt layer of 50~nm thickness is deposited in-situ on both entry and exit openings of the aperture which reduces the probability of forward scattered ions reaching the substrate. The resulting channel length of the aperture in the ion direction is sub-500~nm. This measure lowers the interaction probability between passing ions and aperture walls, and consequently shallow-angle scattering effects. They can be caused by the intrinsic ion beam divergence of about 9~mrad along the aperture tunnel axis. The aperture position relative to the probe tip apex of the cantilever is measured from SEM imaging at the conclusion of the fabrication process. This lateral offset allows AFM topographic images to be precisely aligned with the implant sites from ions passing through the aperture.\\
\textbf{Ion beamline and detection experiments:} The raw P$^+$-ion beam is generated in a BIS DCIS-100 DC plasma filament using a gas intermix consisting of 5\% PF$_5$ diluted in 95\% Argon. A BIS 600-B Wien filter selects the ion species with a beam divergence of typically 9~mrad and about 10~nA beam current. The vacuum pressure inside the ion source chamber is $1\times10^{-6}$~Torr and about $5\times10^{-8}$~Torr in the beam line during operation. A commercial double V-slit configuration made of tantalum membranes and attached to micro-meter screws is employed for beam current adjustment. Further ion beam purification and removal of scattered vacuum background atoms is realised via an NEC 90$^\circ$ electrostatic spherical dipole analyser located at the target chamber entry. A Tungsten membrane of 25~$\mu$m thickness and 20~$\mu$m aperture diameter pre-collimates the ion beam onto the AFM nanostencil aperture. Precision ion current adjustments to $80$ and $150$~ions/s, respectively, are done on a sacrificial detector construction site by monitoring the ion signal rate.\\
A long working distance optical microscope provides a top view on the substrate via a  45$^{\circ}$ mirror with integrated ion aperture. It is used for coarse alignment between AFM nanostencil and ion beam spot as well as the detector relative to the nanostencil. The optical resolution is approximately 2~$\mu$m.\\
\textbf{Theory:} In order to produce a functional qubit via deterministic ion implantation, the following requirements have to be fulfilled: (i) on each location exactly one ion is implanted with a yield defined to be  $Y_\mathrm{DetIon}$; (ii) the final location of the ion in the matrix is compatible with the tolerances of the qubit architecture, which is constrained by the ion straggling and related yield $Y_\mathrm{QCon}$; (iii) the implanted ion is successfully activated upon thermal anneal with the yield $Y_\mathrm{Act}$. The overall yield $Y$ to form a functional donor-qubit then is $Y=Y_\mathrm{DetIon}\cdot Y_\mathrm{QCon}\cdot Y_\mathrm{Act}$.\\
In the following, $Y_\mathrm{DetIon}$ is addressed for the experimental approach presented here and its derivation may differ for other approaches to deterministic ion implantation. In order to successfully record the implantation of an ion, the signal induced by the ion must exceed the data acquisition threshold $q_\mathrm{t0}$. With $F_{\mathrm{Neg}}(q_{\mathrm{t0}})/r_\mathrm{Ion}$ as the fraction of ions that create signals below $q_\mathrm{t0}$, the yield for $s(q)>q_\mathrm{t0}$ is $Y_{s>q_\mathrm{t0}}(q_\mathrm{t0})=1-F_{\mathrm{Neg}}(q_{\mathrm{t0}})/r_\mathrm{Ion}$. Furthermore, the signal recorded above the threshold $q_\mathrm{t0}$ shall not be due to noise events that are erroneously interpreted as an ion implant event. The probability that the signal above $q_\mathrm{t0}$ is related to noise is given as the fraction $F_{\mathrm{Pos}}(q_{\mathrm{t0}})/(F_{\mathrm{Pos}}(q_{\mathrm{t0}})+r_\mathrm{Ion})$. Hence, the yield
$Y_\mathrm{NoNoise}(q_\mathrm{t0})$ that the signal above $q_\mathrm{t0}$ is not due to noise is $Y_\mathrm{NoNoise}(q_\mathrm{t0})=1-F_{\mathrm{Pos}}(q_{\mathrm{t0}})/(F_{\mathrm{Pos}}(q_{\mathrm{t0}})+r_\mathrm{Ion})=r_\mathrm{Ion}/(F_{\mathrm{Pos}}(q_{\mathrm{t0}})+r_\mathrm{Ion})$. The ion beam from the nanostencil dwells on an implant site until a signal above $q_{\mathrm{t0}}$ is recorded. Then the beam is blanked off to protect the implant site from further ion implants and for the nanostencil to re-position to the next implant site. The beam blanker duty cycle time depends on parameters including the signal rise time in the pre-amplifier and related blanker trigger electronics as well as the charging time of the beam electrostatic deflector plates. Realistically achievable blanker times are $\tau\approx100$~ns. Since the ion source delivers the ions stochastically in time, there is a probability $P_\mathrm{DI}$ that a second ion is implanted, creating an unwanted double implant. The probability $P_\mathrm{DI}(\tau, r_\mathrm{Ion})$ is given by $\mathrm{e}^{-\tau\,r_\mathrm{Ion}}$. Hence, the yield that no double implant occurs is 1-$P_\mathrm{DI}$. The yield of successful deterministic ion implants is then given by $Y_\mathrm{DetIon}(q_\mathrm{t0},\tau)=Y_{s>q_\mathrm{t0}}(q_\mathrm{t0})\cdot Y_\mathrm{NoNoise}(q_\mathrm{t0})\cdot (1-P_\mathrm{DI}(\tau, r_\mathrm{Ion}))$. Due to the low ion rate on the order of $~1\,\mathrm{s}^{-1}$ and all ions being implanted in one site (only one blanking cycle at the start and end of experiment) for the experiment presented above, the double implant probability is of the order of $P_\mathrm{DI}\sim 10^{-7}$ and has therefore been neglected in equation \ref{eqn:Xi}, which then reduces to $Y_\mathrm{DetIon}(q_\mathrm{t0})\approx\varXi(q_\mathrm{t0})=Y_{s>q_\mathrm{t0}}(q_\mathrm{t0})\cdot Y_\mathrm{NoNoise}(q_\mathrm{t0})$.\\
\textbf{Simulations:} The simulation of the signal spectrum consists of two steps. At first, the TRIM code \cite{Ziegler1985} is applied to treat the P$^+$-ion transmission through the thin amorphous silicon oxide layer representing the surface gate oxide on our devices. From this it is possible to simulate the energies and directions of a population of P-ions transmitted through the oxide layer and of the recoiled Si and O atoms which are directed into the underlying (100) Si. In the second step the code Crystal-TRIM \cite{Posselt2003, Posselt2006, Pilz1998} is employed to obtain the electronic energy loss per P-ion in the population within the underlying Si. This quantity corresponds to the signal measured by the detector. Finally, detector noise and Fano statistics of e-h pair generation were taken into account and allow the direct comparison with the experimental signal spectrum (see Fig.~\ref{fig:Fig4}a). Crystal-TRIM simulates the trajectories of energetic projectiles (in the present case: P, Si and O) in single crystal Si and can therefore treat channeling effects that cause larger values of electronic energy loss per ion than in amorphous Si. Like TRIM, Crystal-TRIM is based on the binary collision approximation, which assumes that the motion of an energetic projectile may be described by a sequence of binary collisions with target atoms. The trajectory of a projectile between two subsequent collisions is approximated by a straight line given by the asymptote to the trajectory of the energetic particle after the first collision. The electronic energy loss occurring during the collision of a projectile with a target atom is described using a semi-empirical expression, depending on an impact-parameter and is similar to the Oen-Robinson model \cite{Oen1976,Posselt2003, Posselt2006}. In the present work, the value 2.8 was chosen for the model parameter $C_\mathrm{el}$ in the case of P, Si, and O projectiles. The value of the second parameter $C_\mathrm{\lambda}$ was set to 0.92 for P and Si, and to 0.88 for O. This parameter describes the average electronic energy loss for random incidence directions of a projectile. Therefore, $C_\mathrm{\lambda}$ determines the energy related to the histogram peak maximum of the electronic energy loss per incident P ion. The shape of the histogram is sensitive to the parameter $C_\mathrm{el}$, which influences the channeling of a projectile. Thermal vibrations of lattice atoms affect projectile trajectories, especially for the motion in channels. In Crystal-TRIM a simple model is used to take into account this effect \cite{Posselt2003}. Only the motion of P, Si, and O projectiles in single-crystalline Si is followed. However, the Si recoils formed in the collision cascades of these projectiles also contribute to electronic energy loss per incident P ion. This contribution is described by the semi-empirical expression of Funsten et al. \cite{Funsten2001, Funsten2004}, which considers non-negligible self-trapping mechanisms due to a high density of low-energy e-h pairs generated closely around the path of energetic ion projectiles. For the purpose of the present work, the Crystal-TRIM code was modified by the introduction of the Funsten model. This semi-empirial approach replaces earlier models that employed the model of Robinson \cite{Robinson1970}. Test calculations showed that the Robinson model is not capable to describe the electronic energy loss of the low-energy Si recoils to be treated in this work.
\bibliography{sample}



\section*{Acknowledgements}
We thank F.E. Hudson and A.S. Dzurak for discussions and support in the establishment of this research project. The research at the University of Melbourne and UNSW was funded by the Australian Research Council Centre of Excellence for Quantum Computation and Communication Technology (Grant No. CE170100012) and the US Army Research Office (Contract No. W911NF-17-1-0200). We acknowledge a grant from the University of Melbourne  Research and Infrastructure Fund (RIF) and use of the facilities of the Australian National Fabrication Facility (ANFF) at the Melbourne Centre for Nanofabrication (MCN) and at UNSW. H.R.F. acknowledges the support of an Australian Government Research Training Program Scholarship.  A.M. Jakob acknowledges an Australia–Germany Joint Research Cooperation Scheme (UA-DAAD) travel scholarship that supported collaboration with partner institutions in Germany. We are grateful to D. McCulloch of the RMIT Microscopy and Microanalysis Facility for use of SEM/FIB and TEM equipment. The views and conclusions contained in this document are those of the authors and should not be interpreted as representing the official policies, either expressed or implied, of the ARO or the US Government. The US Government is authorized to reproduce and distribute reprints for government purposes notwithstanding any copyright notation herein.

\section*{Author contributions statement}
A.M. Jakob designed and constructed the detector preamplifier electronics and the integration of the ion beam line with the AFM nanostencil scanner. A.M. Jakob conceived and conducted the deterministic implantation experiments with assistance from S.G. Robson. A.M. Jakob, V. Schmitt, V. Mourik and B.C. Johnson developed single ion detectors. V. Schmitt, V. Mourik and H.R. Firgau fabricated single ion detectors. S.G. Robson and B.C. Johnson conducted DIT/FOC as well as C-V/I-V analysis for iterative detector performance improvement. S.G. Robson performed FIB ion-aperture milling and SEM imaging, with assistance from E. Mayes. A.M. Jakob and D. Spemann developed theoretical groundwork on the detection confidence. M. Posselt developed the Crystal-TRIM code. A.M. Jakob and M. Posselt conducted Crystal-TRIM computations and analysis. J.C. McCallum contributed to the design of the ion implantation experiments. A. Morello provided the design of the flip-flop qubit architecture, the design constraints on the fabrication of the detectors, and supervised the research at UNSW. D.N. Jamieson conceived the deterministic ion implantation process and the implementation of the apparatus. The manuscript was written by A.M. Jakob and D.N. Jamieson with contributions from the co-authors.

\section*{Additional information}
The authors declare no competing interests.



\end{document}